# Remote Surface Optical Phonon Scattering in Ferroelectric $Ba_{0.6}Sr_{0.4}TiO_3$ Gated Graphene

Hanying Chen[1,2], Tianlin Li[1,2], Yifei Hao[1,2], Anil Rajapitamahuni[1,2], Zhiyong Xiao[1,2], Stefan Schoeche[3], Mathias Schubert[2,3], and Xia Hong[1,2,a]

[1] Department of Physics and Astronomy, University of Nebraska-Lincoln, Lincoln, Nebraska, 68588-0299, USA

[2] Nebraska Center for Materials and Nanoscience, University of Nebraska-Lincoln, Lincoln, Nebraska, 68588-0299, USA

[3] Department of Electrical and Computer Engineering, University of Nebraska-Lincoln, Lincoln, Nebraska, 68588-0511, USA

[a] Correspondence Email: xia.hong@unl.edu

**Abstract**

We report the effect of remote surface optical (RSO) phonon scattering on carrier mobility in monolayer graphene gated by ferroelectric oxide. We fabricate monolayer graphene transistors back-gated by epitaxial (001) $Ba_{0.6}Sr_{0.4}TiO_3$ films, with field effect mobility up to 23,000 $cm^2\,V^{-1}\,s^{-1}$ achieved. Switching the ferroelectric polarization induces nonvolatile modulation of resistance and quantum Hall effect in graphene at low temperatures. Ellipsometry spectroscopy studies reveal four pairs of optical phonon modes in $Ba_{0.6}Sr_{0.4}TiO_3$, from which we extract the RSO phonon frequencies. The temperature dependence of resistivity in graphene can be well accounted for by considering the scattering from the intrinsic longitudinal acoustic phonon and the RSO phonon, with the latter dominated by the mode at 35.8 meV. Our study reveals the room temperature mobility limit of ferroelectric-gated graphene transistors imposed by RSO phonon scattering.

## I. Introduction

Leveraging its high mobility, superb mechanical strength, and optical transparency, extensive research has been carried out on graphene based field effect transistor (FET) devices[1, 2] for developing radio-frequency transistors,[3, 4] nonvolatile memory,[5] flexible electronics,[6] and optoelectronics.[7, 8] As the electronic properties of graphene are highly susceptible to the interfacial dielectric environment due to its two-dimensional (2D) nature, the choice of the gate and substrate materials for graphene FETs can have a significant impact on the device performance.[9] For example, the gate can be a major source of charged impurities[10] as well as providing dielectric screening.[11-13] The interfacial charge dynamics can induce undesired switching hysteresis and compromise the device retention.[9, 14] One important phenomenon is the remote surface optical (RSO) phonon from the dielectric layer, which is the major mechanism that limits room temperature mobility[15-17] and saturation current[9, 18, 19] in the graphene channel.

Recently, ferroelectric/2D van der Waals heterostructure has emerged as a promising platform for developing high performance logic, memory, and optical applications.[14, 20] Ferroelectrics possess nonvolatile switchable polarization with high doping capacity,[21] and exhibit second harmonic generation[22] as well as negative capacitance effect,[23] which can be utilized to design novel functionalities in the 2D devices, such as electron collimation,[24] nonlinear optical filtering.[25], steep slope switching,[26] and neuromorphic computing.[27] The widely explored ferroelectric systems include the ferroelectric oxides,[6, 8, 28-30] polymers,[5, 31, 32] and 2D semiconductors.[20] Among them, the ferroelectric perovskite oxides have the distinct advantages of large bandgap, large polarization, high-$\kappa$ dielectric constant, high endurance, low coercive field, and fast switching time.[14] For electronic applications, it is important to understand the effect of interfacial ferroelectric layer on the channel mobility. The high-$\kappa$ nature of ferroelectric oxides implies the presence of soft optical phonon modes, which can significantly affect the mobility of the 2D channel at room temperature.[12, 33] This effect is especially prominent in monolayer graphene (1LG) due to its linear dispersion.[9, 34] Despite the rapid progress in developing ferroelectric/graphene-based applications, the effect of RSO phonon on the electronic and thermal transport of the device has not been examined to date.

In this work, we report a comprehensive magnetotransport study of graphene FETs back-gated by ferroelectric $Ba_{0.6}Sr_{0.4}TiO_3$ (BSTO) thin films. The devices exhibit high field effect mobility up to 23,000 $cm^2$ $V^{-1}$ $s^{-1}$ and nonvolatile field effect modulation of quantum Hall effect. Four RSO phonon modes have been identified by considering the optical phonon modes in BSTO and the dielectric boundary condition. The temperature dependence of resistivity in graphene points to the dominant effect of the 35.8 meV RSO phonon at high temperature. Our study reveals the room temperature mobility limit in graphene imposed by ferroelectric oxide gate, providing important material information for designing high performance ferroelectric/graphene FETs for nanoelectronic applications.

**II. Sample Preparation and Characterization**

We work with epitaxial 100 nm and 300 nm (001) BSTO thin films deposited on Nb-doped $SrTiO_3$ (Nb:STO) substrates using off-axis RF magnetron sputtering. The films are deposited in 25 mTorr process gas of Ar and $O_2$ (ratio 2:1) at 600 °C. X-ray diffraction (XRD) studies show that these films are single crystalline with *c*-axis lattice constant of 3.99 Å (Fig. 1(a)). Atomic force microscopy (AFM) images reveal smooth surface morphology with a typical root-mean-square roughness of 3-4 Å (Fig. 1(a) inset). Thin graphite flakes (kish graphite from Sigma-Aldrich®) are mechanically exfoliated on BSTO. The monolayer flakes are identified optically and characterized via Raman spectroscopy. Figure 1(b) compares the Raman spectra of graphene on BSTO and bare

BSTO. There is no appreciable D band observed in the 1LG sample, indicating a lack of atomic defects. From the Lorentz fits to each peak, we deduce a large 2D to G band intensity ratio $I(2D)/I(G) = 3.7$, which is comparable with those observed in high mobility graphene sandwiched between SiO$_2$ substrates and HfO$_2$ top-layers,[17] confirming the high quality of our samples.[35] Selected flakes are fabricated into Hall bar devices via electron-beam lithography followed by electron-beam evaporation of 5 nm Cr/25 nm Au as the electrodes (Fig. 1(c)). The conductive Nb:STO substrate serves as the back-gate electrode for the field effect studies (Fig. 1(d)). We perform variable temperature magnetotransport measurements in a Quantum Design PPMS using standard lock-in technique (SR830) at an excitation current of 50 nA. The results are based on three graphene samples (denoted as Devices A, B, and C).

The BSTO-gated graphene FETs show high mobility compared with those gated by SiO$_2$. Figure 1(e) shows the Shubnikov-de Haas oscillation of the longitudinal resistivity $\rho_{xx}$ taken on Device A at 10 K and the back-gate voltage of $V_g$ = -1 V. The oscillation period corresponds to a charge density of $n = 5.36 \times 10^{12}$ cm$^{-2}$. The oscillatory amplitude $\delta\rho_{xx}$ is given by:[10]

$$\frac{\delta\rho_{xx}}{\rho_0} = 4\gamma_{th}\exp\left(-\frac{\pi}{\omega_c \tau_q}\right), \tag{1}$$

where $\rho_0 = 157$ Ω is the zero-field resistivity, $\gamma_{th} = \frac{2\pi^2 k_B T/\hbar\omega_c}{\sinh(2\pi^2 k_B T/\hbar\omega_c)}$ is the thermal factor, $\omega_c = eB/m^*$ is the cyclotron frequency with $e$ the elementary charge, and $\tau_q$ is the quantum scattering time. Here $m^* = \hbar\sqrt{\pi n}/v_F$ is the effective mass in monolayer graphene, with $v_F \sim 10^6$ m s$^{-1}$ the Fermi velocity. Fitting $\delta\rho_{xx}/\gamma_{th}$ vs. $1/B$ reveals $\tau_q \sim 33$ fs (Fig. 1(f)), while the extracted Hall mobility $\mu_{Hall} = \frac{1}{ne\rho_0} = 7{,}420$ cm$^2$ V$^{-1}$ s$^{-1}$, corresponding to a transport scattering time of $\tau_t = \frac{m^* \mu_{Hall}}{e} = 200$ fs. These results are comparable with previous reports for graphene on BSTO[29] and SiO$_2$.[10] The large ratio between the transport and quantum scattering times $\frac{\tau_t}{\tau_q}$ ~6.1 indicates the mobility is dominated by small-angle scattering events, *e.g.*, from charged impurities residing within the BSTO substrate.[10] This ratio is larger than those reported for graphene gated by HfO$_2$ and SiO$_2$[9] and agrees well with the theoretical value for long-ranged scatterers considering the dielectric screening of BSTO.[14]

Figure 2(a) shows $\rho_{xx}(V_g)$ of Device A measured at 2 K. We observe a hysteresis between the up-sweep and down-sweep curves, which corresponds to the ferroelectric polarization switching.[14] The Dirac points locate at $V_g = 2.4$ V for the up-sweep state and $V_g = 0.9$ V for the down-sweep state. To determine the carrier density *n*, we characterize the Hall resistivity $\rho_{xy}(B)$ (Fig. 2b inset) at different back-gate voltages and deduce the Hall coefficient $R_H = \rho_{xy}/B$, with $n = 1/eR_H$. We denote the carrier density in the electron and hole doped region as $n_e$ and $n_h$, respectively. Figure 2(b) plots $1/eR_H$ *vs.* $V_g$, where we identify three distinct regions for the up-sweep branch associated with different polarization states of BSTO. When BSTO is uniformly polarized in the $P_{up}$ (region III) and $P_{down}$ (region I) states, it behaves as a normal high-$\kappa$ dielectric. In the intermediate $V_g$-range (region II), we observe a gradual change of carrier density with a steeper slope in $n(V_g)$. As BSTO is a relaxor, the polarization switching is associated with the alignment of polar nanoregions,[36] in contrast to the abrupt switching in canonical ferroelectrics such as Pb(Zr,Ti)O$_3$. The induced polarization corresponds to the enhanced doping efficiency. Similar behaviors are also observed in the down-sweep branch. Around the Dirac point, we fit the conductivity by $\sigma(n) = \left[\left(n_{e,h}e\mu_{FE}\right)^{-1} + \rho_{short}\right]^{-1}$, where $\mu_{FE}$ is the field effect mobility and $\rho_{short}$ is the resistivity due to short-ranged scatterers. Figure 2(a) inset shows the fits to the up-sweep branch, which yields $\mu_{FE}$ of 4,700 cm$^2$ V$^{-1}$ s$^{-1}$ for holes and 23,000 cm$^2$ V$^{-1}$ s$^{-1}$ for electrons. For the down-sweep branch, we extract $\mu_{FE}$ of 2,700 cm$^2$ V$^{-1}$ s$^{-1}$ for holes and 10,300 cm$^2$ V$^{-1}$ s$^{-1}$ for electrons. Figure 2(c) shows $\rho_{xx}(V_g)$ of this device at 8.9 T, which exhibits well-developed quantum Hall states in both branches. The filling factor sequence corresponds to 4(*n* + ½), which is the signature behavior of monolayer graphene.

## III. Temperature dependence of resistivity

To investigate the temperature dependence of resistivity in BSTO-gated graphene, we conduct Hall measurements at different temperatures to convert $V_g$ to *n*. Here we choose to work with the hole-doped region for the up-sweep branch, where we have access to the largest density range. Figure 3(a) shows the *n* vs. $V_g$ at 80 K. BSTO exhibits a linear $n(V_g)$ relation, $n = -\alpha V_g$, with $\alpha$ the gating efficiency. We extract the dielectric constant of BSTO $\epsilon_{BSTO} = \alpha e d/\epsilon_0$, where *d* is the BSTO film thickness, and $\epsilon_0$ is the vacuum permittivity. The temperature dependence of $\epsilon_{BSTO}$ (Fig. 3(b)) is consistent with previous report.[29] We then shift the charge neutral point at different temperatures to $V_g' = 0\ V$, and convert the resulting $\rho_{xx}(V_g')$ (Fig. 3(c) inset) to $\rho_{xx}(n)$ based on the gating efficiency (Fig. 3(c)). At low *n*, $\rho_{xx}$ increases with decreasing temperature, which can be attributed to thermally activated charge carriers in the electron-hole puddle region.[37] At high doping level, the sample exhibits metallic *T*-dependence associated with phonon scattering.[15, 17]

In graphene, the temperature dependence of resistivity can be modeled as:[15]

$$\rho(T, n) = \rho_0(n) + \rho_{LA}(T, n) + \rho_{RSO}(T, n). \tag{2}$$

Here $\rho_0(n)$ is the residual resistivity due to impurity scattering, which is temperature-independent; $\rho_{LA}(T, n)$ is associated with the longitudinal acoustic (LA) phonon scattering intrinsic to graphene; $\rho_{RSO}(T, n)$ is associated with RSO phonon scattering from the interfacial dielectric layer.

### a. Effect of LA phonon

In the nondegenerate equipartition acoustic phonon system,[38] the LA phonon contribution depends linearly on temperature:[15, 28]

$$\rho_{LA}(T,n) = \frac{\pi^2 D_A^2}{2h^2 \rho_m v_{ph}^2 v_F^2} \frac{h}{e^2} k_B T, \quad (3)$$

where $D_A$ is the acoustic deformation potential, $\rho_m = 7.6 \times 10^{-7}$ kg m$^{-2}$ is the areal mass density of graphene, $v_{ph} = 2.1 \times 10^4$ m s$^{-1}$ is the sound velocity for LA phonons in graphene. Equation 3 can well describe the low temperatures data of our samples, where $\rho_{xx}$ exhibits a linearly $T$-dependence that is independent of $n$. From the slopes of low temperature $\rho_{xx}(T)$ for Devices A-C, we obtain $D_A = 20 \pm 6$ eV, which is within the range of previous reports (10 - 30 eV).[9]

### b. Effect of RSO phonon

With increasing temperature, $\rho(T)$ becomes nonlinear and highly dependent on $n$, which can be attributed to the onset of RSO phonon contribution $\rho_{RSO}(T, n)$. The RSO phonon effect has previously been studied in graphene interfaced with SiO$_2$,[15] Al$_2$O$_3$,[16] and HfO$_2$.[17] Under the relaxation time approximation,[12, 34, 39] $\rho_{RSO}(T, n)$ can be expressed as the sum of independent contributions from individual RSO phonon modes:

$$\rho_{RSO}(T,n) = \sum_i \rho_{RSO}^{(i)}(T,n), \quad (4)$$

where $\rho_{RSO}^{(i)}(T,n) = \int dk dq A(\mathbf{k},\mathbf{q}) \frac{g_i}{e^{\hbar\omega_i/k_B T}-1}$ is the contribution from $i$-th RSO phonon mode $\omega_i$. Here $A(\mathbf{k},\mathbf{q})$ is the matrix element for scattering between electron ($\mathbf{k}$) and phonon ($\mathbf{q}$) states; $g_i = \hbar\omega_i \left(\frac{1}{\epsilon_i+1} - \frac{1}{\epsilon_{i-1}+1}\right)$ is the corresponding electron-phonon coupling strength,[17] with $\epsilon_i$ the $i$-th intermediate dielectric constant depending on the frequency dependent dielectric function $\epsilon(\omega)$ of BSTO. At $k_B T \ll E_F$, we can assume $A(\mathbf{k},\mathbf{q})$ and $g_i$ are temperature-independent and thus decouple the density and temperature dependences of resistivity as $\rho_{RSO}^{(i)}(T,n) = C(n) \frac{\tilde{g}_i}{e^{\hbar\omega_i/k_BT}-1}$, where $C(n)$ captures the density dependence of resistivity and $\tilde{g}_i$ is the unitless magnitude of $g_i$.

We first consider the temperature dependence of $\rho_{RSO}^{(i)}$. Phenomenologically, $\epsilon(\omega)$ can be expressed by the generalized Lyddane-Sachs-Teller (LST) relation,

$$\epsilon(\omega) = \epsilon_\infty \prod_i \frac{\omega_{LO}^{(i)\,2} - \omega^2 - i\omega\gamma_{LO}^{(i)}}{\omega_{TO}^{(i)\,2} - \omega^2 - i\omega\gamma_{TO}^{(i)}}, \qquad (5)$$

where $\epsilon_\infty$ is the optical permittivity, $\omega_{TO}^{(i)}$ and $\omega_{LO}^{(i)}$ are the frequencies of the $i$-th transverse optical (TO) and longitudinal optical (LO) phonon modes, respectively. The independent broadening parameters, $\gamma_{TO}^{(i)}$ and $\gamma_{LO}^{(i)}$, account for anharmonic lattice interactions. From Eq. 5, in the case of zero phonon broadening, the dielectric function diverges to infinity at the TO modes and approaches zero at the LO modes. The intermediate dielectric constants $\epsilon_i$ are defined by rewriting the real part of LST relation (Eq. 5) into the form of lossless Lorentzian oscillator approximation,

$$\epsilon(\omega) = \epsilon_\infty + \sum_i (\epsilon_{i-1} - \epsilon_i) \frac{\omega_{TO}^{(i)\,2}}{\omega_{TO}^{(i)\,2} - \omega^2}, \qquad (6)$$

where $\epsilon_{i_{max}}$ is defined as $\epsilon_{i_{max}} = \epsilon_\infty$. From the deduced $\epsilon_i$, we can calculate the electron-phonon coupling strength $g_i$. For graphene sandwiched between BSTO and vacuum, the RSO modes can be determined by matching the dielectric boundary condition $\epsilon(\omega_i) + 1 = 0$.[33, 39]

To obtain $\omega_i$ and $g_i$, we carry out spectroscopic ellipsometry to study the frequencies of the optical phonon modes and dielectric properties of BSTO. Two ellipsometer apparatuses are used for measurements performed at room temperature. A commercial variable angle of incidence spectroscopic ellipsometer (IR-VASE Mark-II; J.A. Woollam Co., Inc.) is employed for the infrared spectral range (650-3000 cm$^{-1}$). An in-house built instrument is used for the far-infrared spectral range (50-650 cm$^{-1}$).[40] Ellipsometry data are obtained in the $\Psi$-$\Delta$ notation, and the measurements are taken at multiple angles of incidence.[41] A bare SrTiO$_3$ substrate is measured in addition for comparison with model calculations using literature values for phonon mode parameters.[42] The ellipsometry data are analyzed for the phonon mode properties of the epitaxial thin film. A three-phase (substrate-film-ambient) model is established, where the film is modeled by the BSTO layer thickness. The far-infrared and infrared dielectric function is modeled using Eq. 5, where the LO phonon mode frequency can be directly obtained from the best-match model analysis. The broadening parameters in the best-match model calculations account for finite absorption losses, or phonon damping, near the TO and LO modes. A detailed discussion of phonon mode analysis can be found in Refs. [43-45]. The BSTO dielectric function is modeled with four phonon mode pairs, $\omega_{TO}^1$-$\omega_{LO}^1$, $\omega_{TO}^2$-$\omega_{LO}^2$, $\omega_{TO}^3$-$\omega_{LO}^3$, $\omega_{TO}^4$-$\omega_{LO}^4$. The calculated $\Psi$ and $\Delta$ data using the above model description are compared against the measured ellipsometry data, while the model parameters are varied until reaching a best match between the calculated and measured data. Parameters for variation are BSTO optical permittivity $\epsilon_\infty$ and film thickness, TO and LO phonon frequencies, and their broadening parameters.

Figure 4(a) shows the frequency dependence of the real part of the complex dielectric function of BSTO $\epsilon_{BSTO}(\omega)$. This function is calculated from Eq. 5 by setting all broadening parameters to zero and using the TO and LO mode parameters obtained from the best-match model calculation using the measured ellipsometry data as target values. In the frequency range of interest, we identify four pairs of TO and LO phonon modes: $\omega_{TO}^1 = 14.4$ meV, $\omega_{LO}^1 = 15.2$ meV; $\omega_{TO}^2 = 17.3$ meV, $\omega_{LO}^2 = 35.8$ meV; $\omega_{TO}^3 = 36.2$ meV, $\omega_{LO}^3 = 58.5$ meV; $\omega_{TO}^4 = 64.0$ meV, $\omega_{LO}^4 = 93.9$ meV. By solving the dielectric boundary condition, we deduce four RSO phonon modes, with each one located between a pair of TO and LO modes: $\omega_1 = 15.1$ meV, $\omega_2 = 35.8$ meV, $\omega_3 = $ 57.9 meV, and $\omega_4 = 86.8$ meV. The corresponding coupling strength is calculated based on the intermediate dielectric constants $\epsilon_i$ deduced from Eq. 6: $g_1 = 0.046$ meV, $g_2 = 4.2$ meV, $g_3 = 0.54$ meV, and $g_4 = 2.1$ meV. Figure 4(b) plots the individual contribution of each RSO phonon scattering to the temperature dependence of resistivity in graphene: $\rho_i(T) \propto \frac{g_i}{e^{\hbar\omega_i/k_BT}-1}$. At $T > 50$ K, it is clear that $\rho(T)$ is dominated by the $\omega_2 = 35.8$ meV mode. The lowest phonon mode $\omega_1 = 15.1$ meV has a negligibly small $g_1$, as revealed by the dielectric function spectrum, so it does not couple strongly to the graphene channel. As a result, even though it has the highest excited phonon population, its contribution to resistivity is insignificant.

Figure 5 shows $\rho_{xx}(T)$ of Device A at various carrier densities beyond the electron-hole puddle region, and the corresponding fitting curves using Eq. 2. We first consider the contributions from all four RSO phonons to $\rho_{RSO}(T,n)$ models. Based on the RSO phonon frequencies $\omega_i$ and their corresponding coupling strength $g_i$, Eq. 4 can be rewritten as:

$$\rho_{RSO}(T,n) = C(n)\left(\frac{0.046}{e^{15.1\,\text{meV}/k_BT}-1} + \frac{4.2}{e^{35.8\,\text{meV}/k_BT}-1} + \frac{0.54}{e^{57.9\,\text{meV}/k_BT}-1} + \frac{2.1}{e^{86.8\,\text{meV}/k_BT}-1}\right) ; \quad (7)$$

As shown in Fig. 5(a), this model well captures the temperature dependence of resistivity at all densities. We then consider fitting $\rho(T)$ using a single effective Bose-Einstein distribution with the dominant phonon frequency of 35.8 meV (Fig. 5(b)):

$$\rho_{RSO}(T,n) = \frac{C'(n)}{e^{35.8\,\text{meV}/k_BT}-1}, \quad (8)$$

where $C'(n)$ is the fitted density dependent resistivity. This model also yields excellent fit to the experimental results for Devices A (Fig. 5(b)), B, and C. These results further confirm the dominating role of the 35.8 meV RSO phonon mode in determining the high temperature resistivity of graphene on BSTO, consistent with the simulation result in Fig. 4(b).

Figure 6(a) shows the density dependence of the coefficient $C(n)$ extracted from these three devices. All samples exhibit a power-law density dependence $C(n) \propto n^{-\beta}$, which can be attributed to the long-range nature of RSO phonon scattering. The exponent $\beta$ ranges from 0.9 to 1.3, consistent with the values reported in graphene interfaced with $SiO_2$[15] and $HfO_2$.[17] This value is higher than the predicted $\beta = 1/2$ for Thomas-Fermi approximation for electron screening.[34] The stronger $n$-dependence has been attributed to the finite-$q$ corrections to the scattering matrix element $|H_{kk'}|^2$.[33, 34]

### c. Modeling of room temperature mobility

Based on the fitting results, we calculate the mobility limit of graphene on BSTO imposed by the LA phonon and various RSO phonon modes using $\mu_i = 1/ne\rho_i(n)$. Figure 6(b) summarizes the mobilities at 300 K from the individual phonon scattering mechanisms using the parameters extracted from Device A. The LA phonon limited mobility has $1/n$ dependence on the density, which gives $\sim 10^5$ cm$^2$ V$^{-1}$ s$^{-1}$ at the doping level of interest. Among all RSO phonons, the 35.8 meV mode is the dominating scattering source at 300 K, yielding $\mu \sim 3 - 4 \times 10^4$ cm$^2$ V$^{-1}$ s$^{-1}$. Combining all phonon contributions, the overall mobility is around $3 \times 10^4$ cm$^2$ V$^{-1}$ s$^{-1}$ and exhibits very weak density dependence. This result is about twice of the room temperature mobility limit for $HfO_2$-gated graphene,[17] despite the significantly higher doping capacity of BSTO.

### IV. Conclusion

In conclusion, we have investigated the effect of RSO phonon scattering in graphene FET back-gated by ferroelectric BSTO. The high temperature resistance is dominated by the 35.8 meV RSO phonon mode, which limits the room temperature mobility of graphene FET to about $3 \times 10^4$ cm$^2$ V$^{-1}$ s$^{-1}$. The BSTO gate provides efficient dielectric screening, high doping capacity, and the promise for local doping control via nanoscale domain patterning, while its impact on the room temperature mobility in graphene FET is highly competitive compared with high-$\kappa$ dielectrics such as $HfO_2$. Our study sheds light on the technological potential of ferroelectric perovskite oxide as the gate material for graphene-based nanoelectronics.

We thank Dawei Li for valuable discussions. This work was primarily supported by the U.S. Department of Energy (DOE), Office of Science, Basic Energy Sciences (BES), under Award No. DE-SC0016153 (graphene FET preparation and characterization). Additional support was provided by NSF through Grant Numbers DMR-1710461 (preparation of BSTO), and EPSCoR RII Track-1: Emergent Quantum Materials and Technologies (EQUATE), Award OIA-2044049 (ellipsometry studies and data modeling). The research was performed, in part, in the Nebraska Nanoscale Facility: National Nanotechnology Coordinated Infrastructure, the Nebraska Center for Materials and Nanoscience, which are supported by NSF ECCS: 2025298, and the Nebraska Research Initiative.

# Figure 1

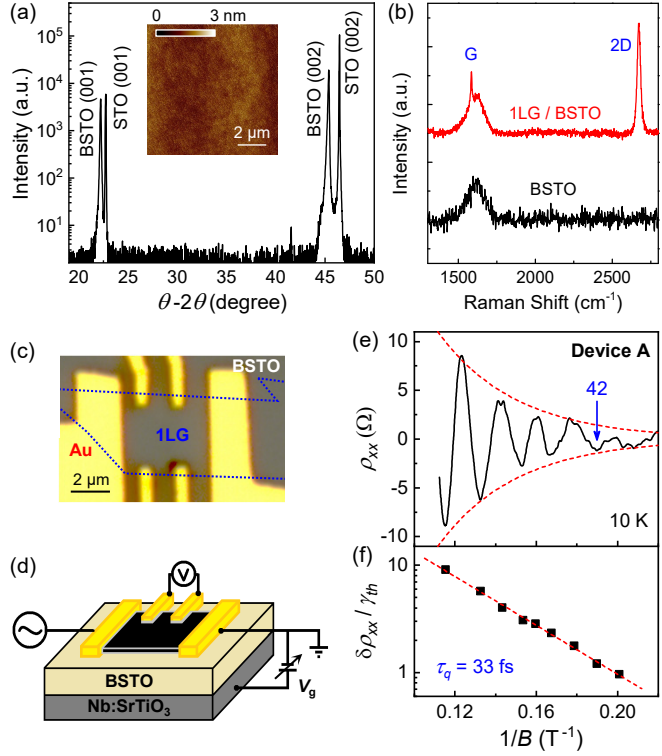

FIG. 1. (a) XRD $\theta$-$2\theta$ scan of a 100 nm BSTO film on Nb:STO. Inset: AFM topography image. (b) Raman spectra of monolayer graphene (1LG) on BSTO normalized to the G peak intensity (red) and bare BSTO (black). (c) Optical image of Device A, with the graphene outlined (dashed line). (d) Device schematic. (e) $\rho_{xx}$ vs. $1/B$ at 10 K with the background resistivity subtracted, and (f) the corresponding semi-log plot of $\delta\rho_{xx}/\gamma_{th}$ vs. $1/B$. The red dashed line is a fit to Eq. (1).

# Figure 2

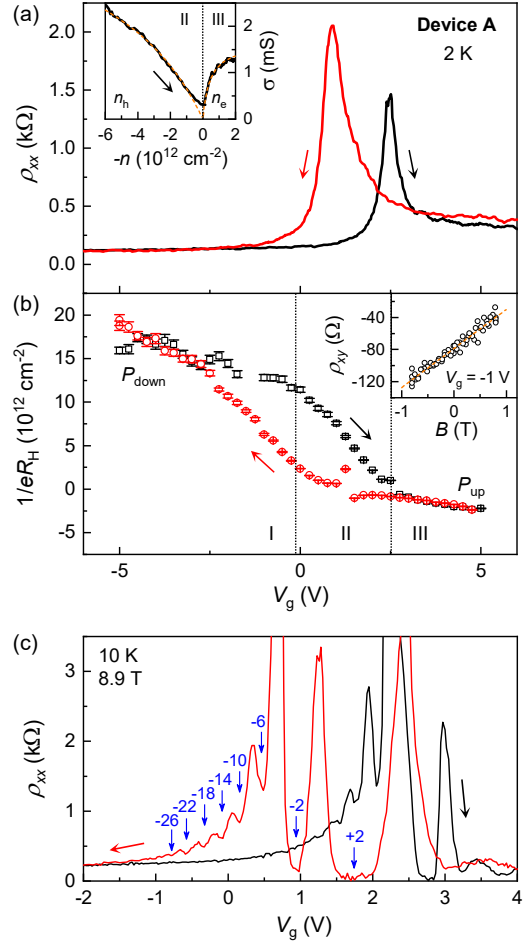

FIG. 2. Magnetotransport studies of Device A. (a) $\rho_{xx}(V_g)$ at 2 K. The arrows label the $V_g$ sweeping direction. Inset: $\sigma$ vs. $-n$ with fits (dashed lines), with $n_h$ and $n_e$ representing the hole and electron doping regions respectively. (b) $1/eR_H$ vs. $V_g$ at 2 K. Inset: $\rho_{xy}$ vs. $B$ at $V_g$ = -1 V for up-sweep with a linear fit (dashed line). (c) $\rho_{xx}(V_g)$ at 10 K and 8.9 T. The filling factors for the down-sweep branch are labeled.

# Figure 3

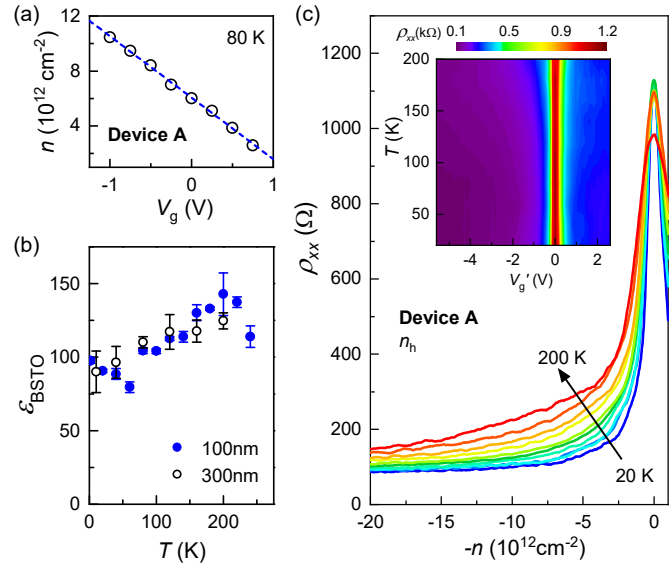

FIG. 3. (a) -$n$ vs. $V_g$ at 80 K with a linear fit (dashed line). (b) $\epsilon_{BSTO}$ vs. $T$ for 100 nm (solid dots) and 300 nm (open circles) BSTO. (c) $\rho_{xx}$ vs. -$n$ at 20 K to 200 K with 20 K intervals. Inset: color map of $\rho_{xx}(V'_g, T)$.

# Figure 4

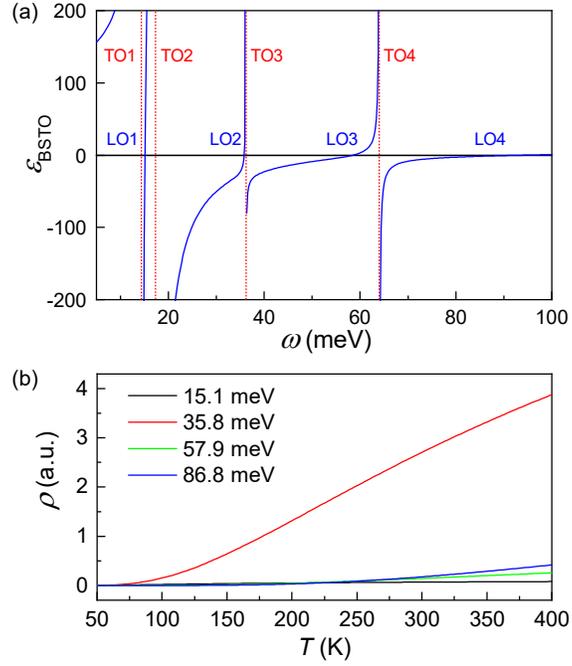

FIG. 4. (a) Real function of $\epsilon(\omega)$ for BSTO simulated based on the LO and TO modes deduced from the ellipsometry data. (b) Simulated $\rho_i(T)$ for the four RSO phonon modes.

# Figure 5

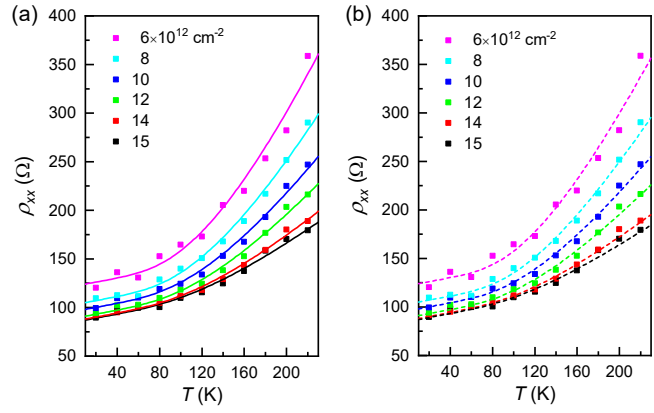

FIG. 5. $\rho_{xx}(T)$ at selected hole densities taken on Device A with fits to Eq. 2. (a) $\rho_{RSO}$ given by Eq. 7 (solid lines). (b) $\rho_{RSO}$ given Eq. 8 (dashed lines).

# Figure 6

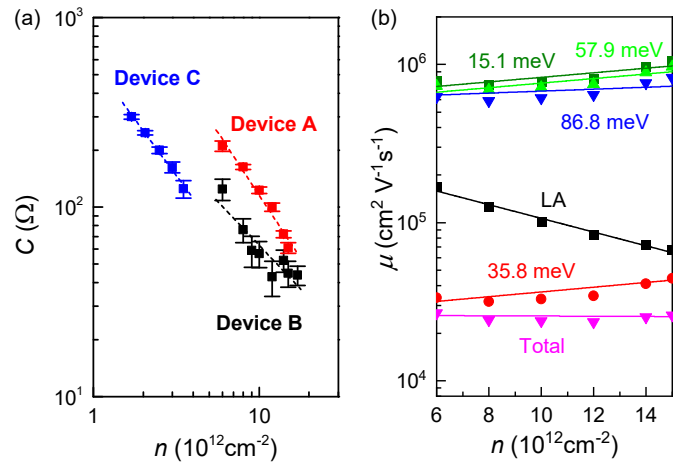

FIG. 6. (a) Resistivity coefficient $C(n)$ vs. carrier density $n$ on Devices A-C with fits to $n^{-\beta}$ (dashed lines). Here $\beta$ = 1.3 (Device A), 0.9 (Device B), and 1.2 (Device C). (b) Mobility limit at 300 K from scattering imposed by LA and RSO phonon modes in Device A.